\documentclass[10pt, a4paper, conference]{IEEEtran}
\usepackage{cite}

\ifCLASSINFOpdf
  \usepackage[pdftex]{graphicx}
\else
  \usepackage[dvips]{graphicx}
\fi
\usepackage{url}

\usepackage{listings}
\usepackage{paralist}
\usepackage{color}
\usepackage{subcaption}
\usepackage{booktabs}
\usepackage{multirow}
\usepackage{amssymb}
\usepackage{amsmath}
\usepackage{hyperref}

\usepackage{flushend}

\usepackage{todonotes}

\definecolor{mygreen}{rgb}{0,0.6,0}
\definecolor{mygray}{rgb}{0.5,0.5,0.5}
\definecolor{mymauve}{rgb}{0.58,0,0.82}
\definecolor{light-gray}{gray}{0.95}

\usepackage{xargs}
\newcommandx{\unsure}[2][1=]{}
\newcommandx{\change}[2][1=]{}
\newcommandx{\info}[2][1=]{}
\newcommandx{\improvement}[2][1=]{}
\newcommandx{\thiswillnotshow}[2][1=]{}

\usepackage{draftwatermark}
\SetWatermarkText{PREPRINT}
\SetWatermarkScale{1}

\begin{document}

\title{The Softwarised Network Data Zoo}


\author{
\IEEEauthorblockN{Manuel Peuster}
\IEEEauthorblockA{Paderborn University\\
manuel.peuster@uni-paderborn.de}
\and
\IEEEauthorblockN{Stefan Schneider}
\IEEEauthorblockA{Paderborn University\\
stefan.schneider@uni-paderborn.de}
\and
\IEEEauthorblockN{Holger Karl}
\IEEEauthorblockA{Paderborn University\\
holger.karl@uni-paderborn.de}
}

%

\maketitle

\begin{abstract}

More and more management and orchestration approaches for (software) networks are based on machine learning paradigms and solutions. These approaches depend not only on their program code to operate properly, but also require enough input data to train their internal models. However, such training data is barely available for the software networking domain and most presented solutions rely on their own, sometimes not even published, data sets. This makes it hard, or even infeasible, to reproduce and compare many of the existing solutions. As a result, it ultimately slows down the adoption of machine learning approaches in softwarised networks.

To this end, we introduce the ``softwarised network data zoo'' (SNDZoo),
an open collection of software networking data sets aiming to
streamline and ease machine learning research in the software
networking domain. We present a general methodology to collect,
archive, and publish those data sets for use by other researches  and,
as an example, eight initial data sets, focusing on the
performance of virtualised network functions.\footnote{\textbf{Full paper will be published in: ``2019 IEEE/IFIP 15th International Conference on Network and Service Management (CNSM), Halifax, Canada.''}}

\end{abstract}

%
\IEEEpeerreviewmaketitle

\section{Introduction}
\label{sec:intro}

The softwarisation of networks is considered as the key enabler for more agile, automated, and autonomous network management concepts, including the use of DevOps paradigms~\cite{karl2016devops}.
The two key technologies to achieve this are software-defined networking (SDN) and network function virtualisation (NFV). Both play an important role as part of an emerging trend, called zero touch network \& service management (ZSM), which aims to remove any manual steps from network management tasks~\cite{etsi.zerotouch.whitepaper2017}.

To automate network management and to realise ZSM, more and more machine learning (ML) and artificial intelligence (AI)-based network management solutions arise which claim to be able to manage and optimise different aspects of our networks~\cite{wang2018mlnetworking}. Many of them focus on automated resource dimensioning for NFV scenarios which try to optimise the amount of resources assigned to each involved virtual network function (VNF). To do so, they predict the upcoming network load as well as the performance a VNF achieves using a given amount of resources \cite{mijumbi2017resourceprediction, sun2018qlearningplacement}.

But ML-based solutions are always data-driven and do not only depend on programming code, which is a huge difference to legacy management and automation approaches. This means that ML approaches can only work efficiently if enough data is available to train and test the involved models, before they are put into production. Even if data is available in some custom environments, e.g., in form of volatile monitoring metrics, we are still missing a publicly available collection of open data sets that can be used to evaluate and compare different ML solutions and algorithms with each other. Such open data sets are a common tool in other communities, like image recognition~\cite{deng2009imagenet} or natural language processing~\cite{maas2011nlpdata}, and accelerated the adoption of ML solutions in those domains.

We argue that the software networking community also needs open data
sets to simplify and streamline ML research and to improve the
reproducibility of new ideas. To this end,  we introduce the
``softwarised network data zoo'' (SNDZoo) -- an open repository to collect, host, and share software networking data sets. The SNDZoo is, to the best of our knowledge, the first effort to build such a central repository for softwarised network data. It  specifically focuses on NFV and SDN performance data sets, collected through performance measurements of real-world network setups. These data sets go beyond existing collections of open networking data sets such as topology data sets or traffic traces~\cite{SNDlib10,topologyzoo2011}.

Our contributions are two-fold. After discussing related work in Section~\ref{sec:rw},  we first introduce the \emph{methodology and workflow} used to collect data sets for the SNDZoo, using our open-source NFV benchmarking platform~\cite{peuster2017chainprofiling}, in Section~\ref{sec:methodology}. Second, we present \emph{eight data sets}, containing millions of performance measurements collected from different real-world VNFs from the security, web, and 5G vertical (Internet of things) areas, in Section~\ref{sec:data set}. We make all data sets  available as part of the SNDZoo project~\cite{web.sndzoo} for use by other researchers.

\section{Related Work}
\label{sec:rw}

As in many other domains, ML recently found its way into the information and communications technology (ICT) sector and networking domain~\cite{wang2018mlnetworking}. The use of ML is especially appealing for network management and optimisation use cases in softwarised networks. It is specifically well suited for NFV/SDN scenarios, which shall be highly automated to allow zero-touch network operations following DevOps methods~\cite{karl2016devops, etsi.zerotouch.whitepaper2017}. Existing work focuses on, e.g., learning and predicting of service metrics, such as response time and frame rate~\cite{stadler2017ml, samani2018mlmetricprediction}, scaling and resource dimensioning~\cite{mijumbi2017resourceprediction} as well as placement decisions~\cite{sun2018qlearningplacement}. But all of this work relies on custom data sets which might not even be publicly available. The available public data sets, such as~\cite{stadler2017ml}, are however not available at a common repository and thus are often hard to find. Our work improves this situation by offering a common repository to host and share data sets focusing on performance measurements of softwarised networks as well as the involved platforms and components. This also complements and goes beyond existing collections of open network data sets mainly focusing on network topology graphs, like~\cite{topologyzoo2011, web.snapnets, SNDlib10}.


To collect the presented data sets, automated solutions to benchmark and profile NFV and SDN scenarios, including our own work~\cite{peuster2016profiling, peuster2017chainprofiling}, can be used~\cite{rosa2017gym, cao2015nfvvital, nam2018probius, khan2018nfv}. Those benchmarking approaches have not only been proposed by academia but have also been in the focus of standardisation bodies~\cite{ietf.rosa-bmwg-vnfbench, etsi.nfv.tst009}. They are complementary to the work presented in this paper. In fact, we make use of these solutions to collect the initial data sets available in the SNDZoo~\cite{web.sndzoo}.

\section{Methodology \& Workflow}
\label{sec:methodology}

Collecting data sets from softwarised network scenarios is more than performing a handful of manual measurements on a testbed running in a lab. In fact, manual measurements should be avoided where possible to be able to (i) quickly and objectively reproduce the measurements, (ii) run a given measurement in a new environment, e.g., outside the lab, and (iii) make use of the fact that software-based networking scenarios can be automatically deployed, allowing to completely remove all manual steps from the process. On top of that, software-based networks usually offer a much higher degree of configuration freedom, e.g., in terms of virtual resources assigned to a VNF, resulting in many different configurations for which measurements can be and need to be performed~\cite{peuster2018tcp}.  

\begin{figure}[t]
	\centering
	\includegraphics[width=1.0\columnwidth]{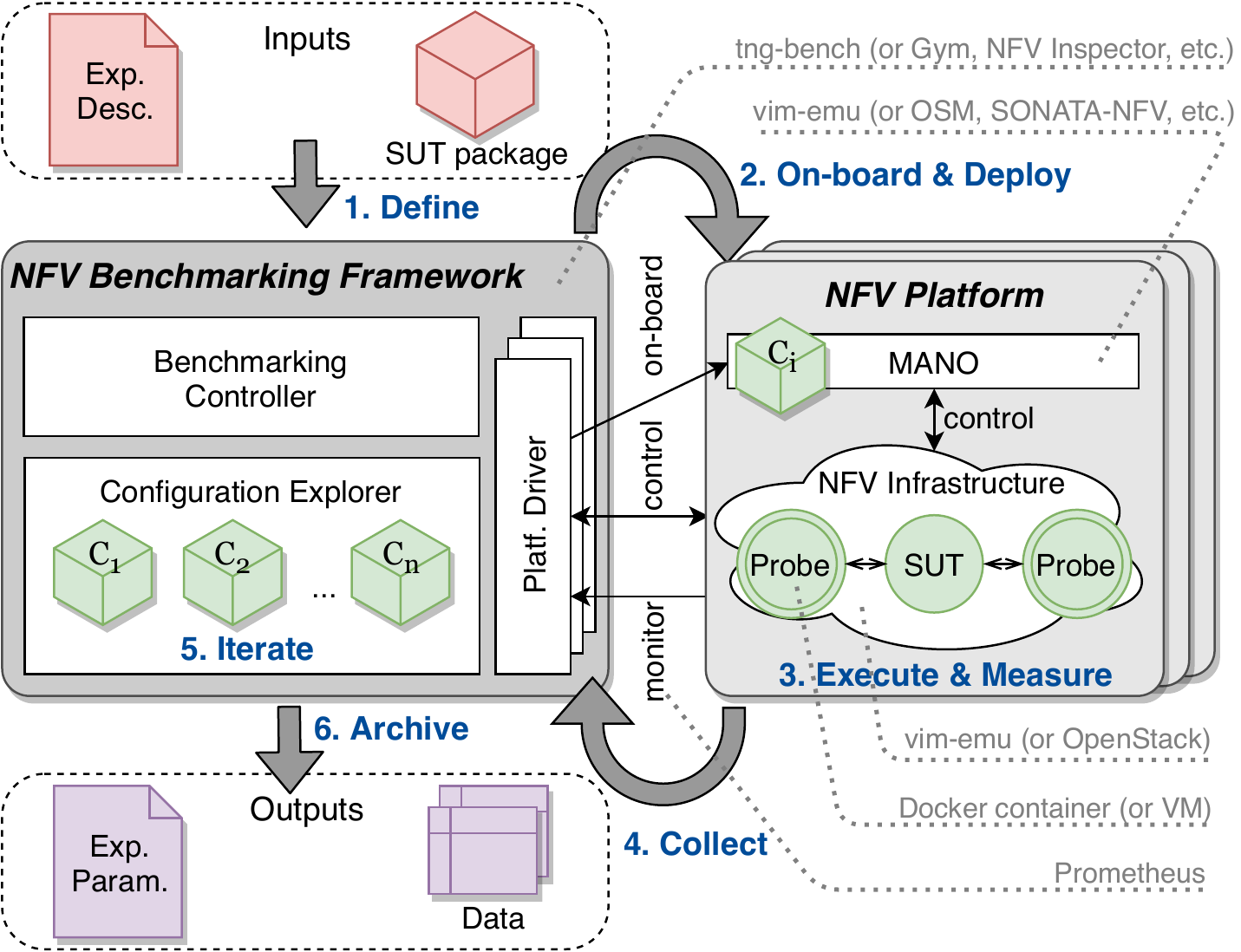}
	\caption{Example data collection setup and workflow. The dotted notes point to possible implementation options.}
	\label{fig:setup}
\end{figure}

We introduce such a fully-automated data collection methodology in Figure~\ref{fig:setup} using an NFV scenario as example. A data collection setup consists of two main components. First, an NFV benchmarking framework that is responsible to control, manage, and automate the measurement and collection process. Second, one or multiple NFV platforms, consisting of management and orchestration~(MANO) layer and NFV infrastructure. They are used to deploy and execute the experiment setups, including the deployment and execution of the system under test~(SUT). This concept is independent of the technical realisation of the different components. It is, for example, possible to use different benchmarking frameworks, such as tng-bench~\cite{peuster2017chainprofiling}, Gym\cite{rosa2017gym}, or NFV Inspector \cite{khan2018nfv}, or multiple NFV platforms, such as vim-emu~\cite{peuster2016medicine, web.osm.vimemu.wiki}, OSM~\cite{web.osm}, or SONATA-NFV~\cite{web.sonata}, to perform measurements in different environments.

Using the setup shown in Figure~\ref{fig:setup}, the following end-to-end workflow is used to collect a full data set, containing measurements for many different configurations of a given SUT. First, the measurement experiment is defined using the description approach offered by the used benchmarking framework~(1). This description contains, e.g., a list of different configurations that should be tested as well as a definition of the probes that should be used. Probes are VNFs that can be deployed together with and connected to the SUT and are used to stimulate the SUT during the experiments. Those stimuli can be, e.g., synthetic traffic generators or replayed traffic traces. The final description as well as the SUT (e.g., given as an ETSI SOL004 VNF package) is then used as input for the framework. The framework then on-boards and deploys the first configuration to be tested on the NFV platform by triggering the MANO system~(2). Once the SUT and the defined probes are instantiated and configured, the framework triggers the execution of the experiment, i.e., activating the probes and measuring the performance achieved by the SUT~(3). When the execution is done (e.g., after a fixed time limit), the SUT and the probes are terminated and the results are collected by the framework~(4), before the next configuration is selected~(5) and a new measurement cycle is started~(2). Finally, after all configurations have been successfully executed, the framework combines all collected results as well as the configuration parameters used for each iteration and archives them in a table-based format~(6).

We implement the presented workflow using our own NFV benchmarking framework, called \emph{tng-bench}, which was initially presented in~\cite{peuster2017chainprofiling} and is now available as actively-developed open-source project~\cite{web.repo.tngbench}. We use vim-meu~\cite{peuster2016medicine} as NFV platform to execute container-based SUTs and probes. 
There are two kinds of metrics collected by tng-bench. First, the experiment metrics that are collected from the probes as well as the SUT at the end of an experiment, e.g., total number of processed packets. Those metrics are captured by collecting log files from the involved containers before termination. Second, we collect time series metrics using the Prometheus time series database~\cite{web.prometheus} controlled by tng-bench. Prometheus periodically fetches all metrics, including the resource usage of the involved containers (using cAdvisor~\cite{web.cadvisor}) as well as SUT-specific metrics as we detail in the following section.


\section{Collecting, Archiving, and Publishing\\the First Data Sets}
\label{sec:data set}

We collected eight initial data sets, focusing on the performance of real-world VNFs under different configurations 
to kick-off the SNDZoo project and to test the presented methodology, workflow, and tools. All initial data sets are automatically collected and can be reproduced using our benchmarking tool. We want to highlight that the scope of the SNDZoo is not limited to performance data sets of single VNFs. We plan to extend our efforts and also include measurements of more complex network services, MANO performance numbers, impact of different management approaches to the performance of real world NFV systems, or even long-term monitoring data into the project. 

\subsection{Experiment Setup}

To collect the data sets, we picked eight VNFs from three different categories as shown in Table~\ref{tab:datasets}: Security (IDS systems), web (load balancers, proxies), and IoT (MQTT brokers). This way, we not only have VNFs that transparently forward the traffic while passively analysing it (IDS systems), but also active VNFs that can modify the traffic (proxies). We also have scenarios (MQTT broker) that can be considered as examples for 5G vertical use cases, such as IoT, smart manufacturing, or industry 4.0~\cite{schneider19industrypilot}.

In the first category (\texttt{SEC01}\,-\,\texttt{SEC03}), we benchmark three IDS VNFs, namely Suricata~4.0~\cite{web.tool.suricata}, Snort~2.9~\cite{web.tool.snort}, and Snort~3.0~\cite{web.tool.snort}. Each IDS is configured as transparent layer~2 bridge and passively monitors the incoming traffic. To stimulate the VNFs we use two publicly available traffic traces with small and big flows that are continuously replayed at maximum speed~\cite{web.traces1}. During the benchmarking experiment, the VNFs are configured with different IDS rule sets, taken from~\cite{web.ids_rules1}, and with different resource assignments, i.e., CPU time (10\,\% to 100\,\%) and memory (256\,MB and 1024\,MB). As a result, 80 different configurations are tested\footnote{There are only 40 configurations in the case of Snort~3.0 because of the smaller rule set available for this version of Snort.} and each configuration is repeated 20 times, resulting in 1,600 experiment runs, each testing a single configuration.

The second category (\texttt{WEB01}\,-\,\texttt{WEB03}) represents web scenarios in which we test an Nginx~1.10.3~\cite{web.tool.nginx} and a HAProxy~1.6.3~\cite{web.tool.haproxy} load balancer VNF as well as a Squid~3.5.12~\cite{web.tool.squid} proxy. The VNFs are placed between a source probe (user requests generated by Apache Bench~\cite{web.tool.apache}) and a target probe (web server running Apache~2.0~\cite{web.tool.apache}). As shown in Table~\ref{tab:datasets}, 80 different configurations are executed, including small and large requests and different resource assignments, i.e., CPU time (10\,\% to 100\,\%) and memory (64\,MB, 128\,MB, 256\,MB, and 512\,MB).

In the third category (\texttt{IOT01}\,-\,\texttt{IOT02}), the MQTT brokers Mosquitto~1.6.2~\cite{web.tool.mosquitto} and Emqx~3.1.0~\cite{web.tool.emqx} are tested using Malaria~\cite{web.tool.malaria}, an MQTT load generator. The broker is placed between two probes running Malaria instances, one acting as publisher and the other acting as subscriber, allowing us to measure the end-to-end delay of MQTT messages. Besides different resource assignments for the broker VNF, Malaria is executed with different configurations, e.g., message sizes between 10 and 1000\,bytes as well as two different MQTT QoS levels (1 and 2). Please note that the full details of all used configurations, versions, and workloads are published along with the data sets~\cite{web.sndzoo}.

\begin{table*}
    \centering
    \caption{Overview of the eight VNF benchmarking data sets initially published in the SNDZoo~\cite{web.sndzoo}}
    \begin{tabular}{lllllp{1.0cm}p{1.0cm}p{1.0cm}p{.9cm}p{1.0cm}p{1.0cm}}
        \toprule
        Name & Category & Class & VNF & Probe/Stimuli & Tested Configurations & Repetitions & Experiment Metrics & Time Series Metrics & Total Exp. Runtime & Total data points\\
        \toprule
		\texttt{SEC01} & Security & IDS Systems & Suricata~\cite{web.tool.suricata} & Traces~\cite{web.traces1} & 80 & 20 & 280 & 157 & 38.1\,h & 7.9M \\ 
		\texttt{SEC02} & &  & Snort 2.9~\cite{web.tool.snort} & Traces~\cite{web.traces1} & 80 & 20 & 280 & 169 & 37.6\,h & 8.6M \\ 
		\texttt{SEC03} & &  & Snort 3.0~\cite{web.tool.snort} & Traces~\cite{web.traces1} & 40 & 20 & 281 & 593 & 19.2\,h & 14.5M \\ 
		\midrule
		\texttt{WEB01} & Web & Load balancers & Nginx~\cite{web.tool.nginx} & AB/Apache~\cite{web.tool.apache} & 80 & 20 & 268 & 43 & 38.3\,h & 2.5M \\ 
		\texttt{WEB02} & & & HAProxy~\cite{web.tool.haproxy} & AB/Apache~\cite{web.tool.apache} & 80 & 20 & 268 & 43 & 39.4\,h & 2.5M \\ 
		\texttt{WEB03} & & Proxys & Squid~\cite{web.tool.squid} & AB/Apache~\cite{web.tool.apache} & 80 & 20 & 268 & 43 & 39.9\,h & 2.5M \\ 
		\midrule
		\texttt{IOT01} & IoT & MQTT Broker & Mosquitto~\cite{web.tool.mosquitto} & Malaria~\cite{web.tool.malaria} & 80 & 20 & 275 & 90 & 40.0\,h & 4.7M \\ 
		\texttt{IOT02} & &  & Emqx~\cite{web.tool.emqx} & Malaria~\cite{web.tool.malaria} & 80 & 20 & 275 & 90 & 79.8\,h & 4.7M \\  
		\bottomrule
    \end{tabular}
    \label{tab:datasets}
\end{table*}

\subsection{Data Collection}

To collect the presented data sets we used the setup presented in Section~\ref{sec:methodology} using vim-emu~\cite{peuster2016medicine} as NFV platform, which is executed on a machine with Intel(R) Xeon(R) W-2145 CPU at 3.70\,GHz CPU, 32\,GB of memory, running Linux 4.4.0-142-generic\footnote{The full hardware and software specifications of the used testbed are available as part of the SNDZoo repositories.}. Vim-emu allows to deploy and control NFV scenarios on a single physical machine using Docker containers. In our experiments, the tested VNFs as well as the probes used to stimulate them are deployed as Docker containers, each of them always pinned to a single physical CPU core to achieve isolation between VNFs and probes~\cite{peuster2017chainprofiling}. 

We collect a large number of different experiment metrics for each tested configuration as well as a large number of time series metrics during experiment execution.
More specifically, we collect 268 to 281 experiment metrics after each experiment and between 43 and 593 time series metrics during each experiment. This results in up to 474,400 collected time series records in data set \texttt{SEC03}\footnote{Considering configurations, repetitions, and collected metrics results in: 40$\cdot$20$\cdot$593=474,400 time series records.}, as shown in Table~\ref{tab:datasets}. Each of these records contains about 30 data points resulting from a collection frequency of 0.5\,Hz and an experiment runtime of 60\,s per configuration. Those numbers highly depend on the involved VNFs and the number of metrics they expose. The used Snort~3.0 VNF, for example, exposes more than 400 metrics that can be collected, e.g., packet counters for different protocol types.

The use of tng-bench as experiment automation framework allows to reproduce all presented experiments. To do so, nothing more is required than two Linux machines on which tng-bench and vim-emu are installed. All involved VNFs can be downloaded from the SNDZoo repositories as pre-defined Docker containers~\cite{web.sndzoo} and used to re-run the experiments. 
However, the absolute performance numbers in those data sets obviously depend on the underlying hardware and will differ for new measurements performed in different environments. But, generic aspects  like trends that can be identified between different VNF configurations will still be visible~\cite{peuster2016profiling}.

\subsection{Resulting Data Sets}

The presented data sets contain between 2.5 and 14.5 million data points and are, for example, usable as training and testing data sets for different kinds of prediction or optimisation algorithms in the NFV domain. The collection process of each data set took between 19.2\,h and 79.8\,h, as shown in Table~\ref{tab:datasets}. Even though the \texttt{SEC03} data set has the shortest runtime, it contains the most data points. The reason for this is that the tested VNF, Snort~3.0, exposes more VNF-specific metrics than any other tested VNF. The measurements to collect \texttt{IOT02} took more time than the others because the used VNF (Emqx) takes much longer to start up and to be ready to process traffic.

\begin{figure}[!ht]
\centering
\begin{subfigure}{1.0\columnwidth}
  \includegraphics[width=1.0\linewidth]{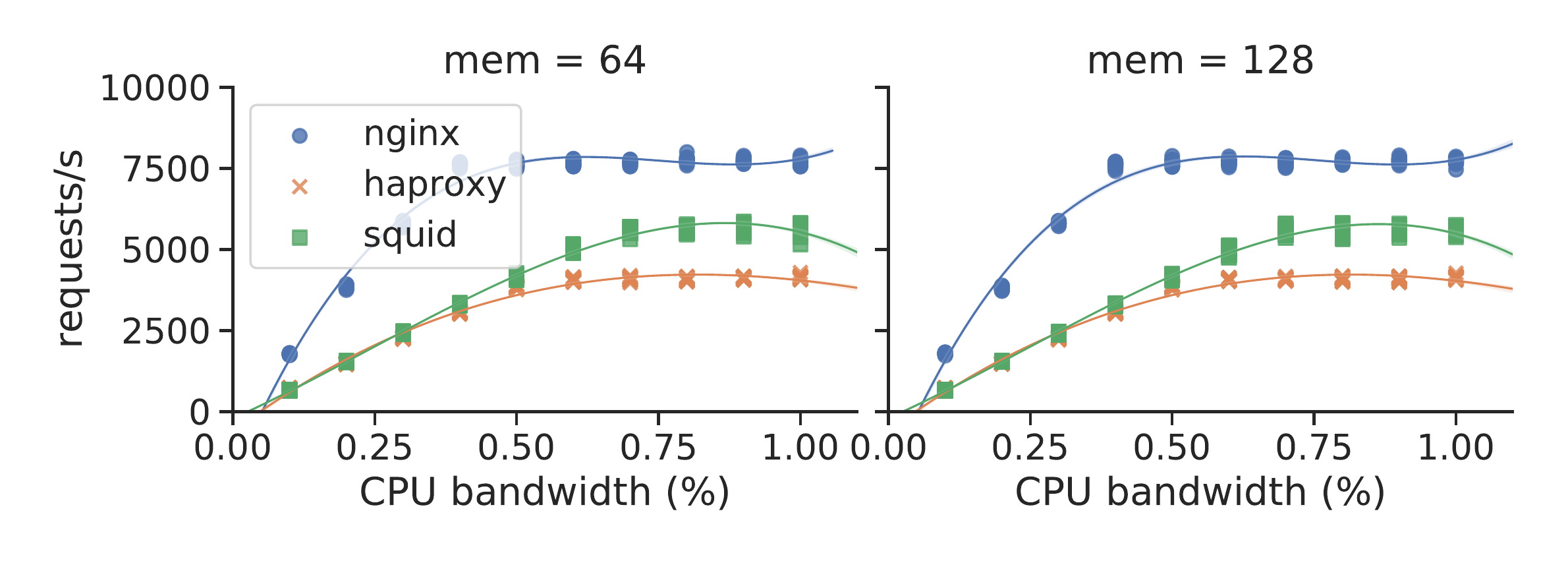}
  \vspace{-0.5cm}
  \label{fig:plot11}
\end{subfigure}
\newline
\begin{subfigure}{1.0\columnwidth}
  \includegraphics[width=1.0\linewidth]{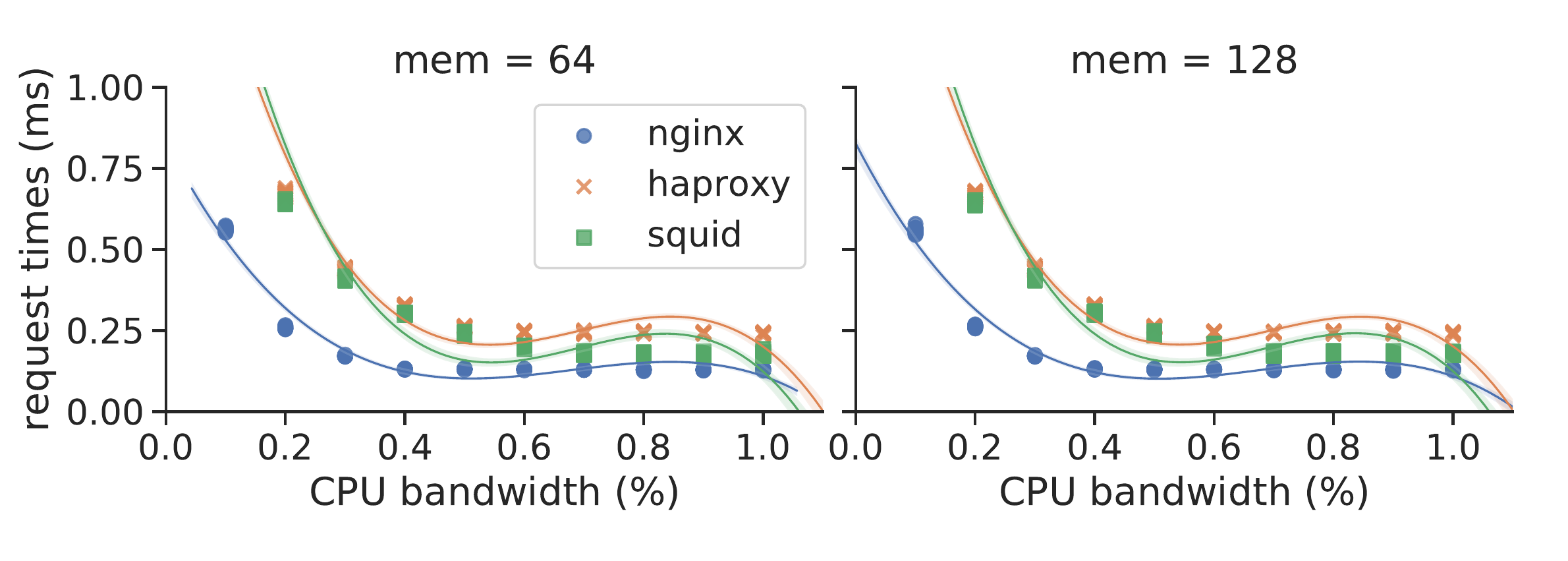}
  \vspace{-0.5cm}
  \label{fig:plot12}
\end{subfigure}

\caption{Example data set visualisations}
\label{fig:plot1}
\end{figure}

Figure~\ref{fig:plot1} visualises a small subset of our data sets to give the reader a brief example of the published data and to show how researchers can make use of it. The figure shows the comparison of two metrics (requests/s and request times) of the \texttt{WEB01}-\texttt{WEB03} data sets for different CPU and memory configurations of the three SUTs (nginx, haproxy, and squid). Each configuration is represented by 20 points (measurement repetitions). The lines indicate polynomial regressions (degree=3) to demonstrate how the data can be used to train performance models mapping different SUT configurations to performance metrics. Those models could, e.g., be used for scaling and placement optimisation algorithms~\cite{draexler2018jasper}.

\subsection{Publishing the Data Sets}

We version all experiment configurations that are used to collect the presented data sets using Git so that we can reproduce old experiments even if they are updated or improved over time. This also allows us to use GitHub as hosting platform to publish and share the contents of the SNDZoo. However, keeping large data sets, containing multiple files with gigabytes of data, within a Git repository is bad practice and results in poor performance. To solve this, we make use of a recently introduced project called Data Version Control (DVC)~\cite{web.tool.dvc}. DVC focuses specifically on versioning ML data sets and allows to store and version large files on external storage solutions such as Amazon S3 while referencing them from a Git repository. Users can then access and download a specific data set by simply running two commands (\texttt{git clone} and \texttt{dvc pull}) on their machine. 

Each data set hosted in the SDNZoo is located in its own Git/DVC repository with a connected Amazon S3 bucket holding the data files (a total of 3.0\,GB). Those repositories not only contain the configurations and resulting data sets but also meta data like the hardware and software specifications of the machines on which the measurements have been performed. Further, they contain license information as well as all raw measurements, time series, and logs produced by tng-bench.
Besides the data set repositories,  SNDZoo also provides repositories containing the sources and descriptions of used VNFs and services. 
All published data sets are indexed on and linked from the SNDZoo website~\cite{web.sndzoo}. They are published under creative commons CC-BY-SA~4.0 license.

\section{Conclusion}
\label{sec:conclusion}

We introduced the SNDZoo project as the first open collection of NFV/SDN performance data sets. The SNDZoo project aims to support the adoption of ML/AI in the software networking community by enabling researchers to work with common data sets simplifying  comparisons and reproduction of results. Along with this paper, we publish eight initial data sets, containing performance measurements collected from a set of real-world security, web, and IoT VNFs.

We plan to continue our efforts and add new data sets over time. Our broader vision for the SNDZoo is to have a large collection of different data sets for a wide variety of use cases, scenarios, and deployments. Not only focusing on NFV and SDN performance measurements but also on measurements of MANO system and management performance. However, this project will depend on community contributions and we invite all community members to participate and use SNDZoo as a platform to host and share their data sets.
Contributors are free to choose how they collect their data sets as long as they ensure that the data sets come with enough information such that the measurements can be reproduced in a fully automated fashion.
More details about the contribution process are available online~\cite{web.sndzoo}.

\section*{Acknowledgments}
\scriptsize{
This work has received funding from the European Union's Horizon 2020 research and innovation programme under grant agreement No. H2020-ICT-2016-2 761493 (5GTANGO), and the German Research Foundation (DFG) within the Collaborative Research Centre ``On-The-Fly Computing" (SFB 901).}

\bibliographystyle{IEEEtran}
\bibliography{IEEEabrv,main}

\end{document}